\documentclass[10pt,superscriptaddress,english,twocolumn,showpacs,
				floatfix,aps]{revtex4-2} 
\usepackage[utf8]{inputenc}
\usepackage{amsmath}
\usepackage{amssymb}
\usepackage{graphicx}
\usepackage{xspace}
\usepackage{soul}
\usepackage{braket}
\usepackage[backref=none,bookmarksnumbered=true,bookmarks=true,
	bookmarksopen=true,colorlinks=true,citecolor=blue,linkcolor=blue,
	anchorcolor=green,urlcolor=blue,unicode=false]{hyperref}

\newcommand{\rj}{\ensuremath{\mathrm{R_J}}\xspace}

\begin{document}

\title{Shot-noise measurements of single-atom junctions}

\author{Idan Tamir}
\email{Idan.Tamir@FU-Berlin.de}
\affiliation{Fachbereich Physik, Freie Universit\"{a}t Berlin, 14195 Berlin, Germany.}
\author{Verena Caspari}
\affiliation{Fachbereich Physik, Freie Universit\"{a}t Berlin, 14195 Berlin, Germany.}
\author{Daniela Rolf}
\affiliation{Fachbereich Physik, Freie Universit\"{a}t Berlin, 14195 Berlin, Germany.}
\author{Christian Lotze}
\affiliation{Fachbereich Physik, Freie Universit\"{a}t Berlin, 14195 Berlin, Germany.}
\author{Katharina J. Franke}
\affiliation{Fachbereich Physik, Freie Universit\"{a}t Berlin, 14195 Berlin, Germany.}

\date{\today}

\begin{abstract} 
Current fluctuations related to the discreteness of charge passing through small constrictions are termed shot noise. This unavoidable noise provides both advantages - being a direct measurement of the transmitted particles' charge, and disadvantages - a main noise source in nanoscale devices operating at low temperature. While better understanding of shot noise is desired, the technical difficulties in measuring it result in relatively few experimental works, especially in single-atom structures. Here we describe a local shot-noise measurement apparatus, and demonstrate successful noise measurements through single-atom junctions. Our apparatus, based on a scanning tunneling microscope operates at liquid helium temperatures. It includes a broadband commercial amplifier mounted in close proximity to the tunnel junction, thus reducing both thermal noise and the input capacitance that limit traditional noise measurements. The full capabilities of the microscope are maintained in the modified system and a quick transition between different measurement modes is possible.      
\end{abstract}

\maketitle
\section{Introduction}

Due to the statistical nature of transmission of particles through non-transparent junctions, current fluctuates in time. These fundamental fluctuations, related to the discreteness of charge-carrying particles, lead to the so-called shot noise in small constrictions \cite{Schottky}. 
It is precisely this sensitivity to the charge of the transmitted particles (or quasi-particles), which entails great potential to the study of electron-electron correlation effects and exotic electronic states (for a review see Ref. \cite{BLANTER2000}). For instance, shot-noise measurements have provided evidence of fractional charges in the quantum Hall effect \cite{Saminadayar1997,Picciotto1998,Bid2009} and Cooper-pair transport in superconductors \cite{jehl2000,Bastiaans2019,Koen2021}. It may further yield information of the scattering processes at Kondo impurities \cite{Meir2002,Sela2006,Zarchin2008,delattre2009,Yamauchi2011,Burtzlaff2015,ferrier2016,Cocklin2019,Mohr2020}.

Hence, shot noise provides valuable information of fundamental physical processes beyond other experimental techniques. Combined with atomic-scale imaging, it could open the door to the characterization of electron correlations in nanoscale materials. Unfortunately, the measurement of shot noise is a challenging and demanding task. Firstly, other noise sources are ever-present in real systems and need to be disentangled. Secondly, the shot-noise signal is very small ($\sim100$ fA/$\sqrt{\text{Hz}}$). As a result, its measurement requires both to minimize other sources of noise while simultaneously amplifying the shot-noise component. 

The main unavoidable sources of noise are thermal \cite{Johnson1928,Nyquist1928} and $1/f$ (also called pink) noise \cite{Johnson1925}. To reduce thermal noise, the junction itself needs to be placed at low temperatures. Yet, this is not sufficient since all electronic components in the setup also contribute to the total thermal noise. It is thus important that measurement-related electronics are also placed in a low temperature environment. The second prerequisite is to measure at high enough frequencies, where the amplitude of the $1/f$ noise can be disregarded, typically well above 1 kHz. At higher frequencies, however, there is a natural cutoff due to the system's finite resistance and capacitance to ground that form an effective low-pass filter. Finally, one has to address 50- (or 60-)Hz noise, radiated from all grid-powered electronic devices and coupled to the measurement setup. This 50-Hz noise is practically unavoidable, but can be reduced by properly grounding the setup (adopting a star-shaped grounding scheme, and providing independent ground), supplying power by using batteries, and connecting isolation transformers to buffer those devices that are not battery powered.

To interpret the shot-noise signal and to draw conclusion on the physical properties of the junction, we briefly describe the fundamental properties of transport through mesoscopic systems \cite{BLANTER2000}. The transmission probability of non-interacting electrons passing through a narrow constriction is expected to follow a Poisson distribution. The associated noise in the current is determined by the total conductance:
$S_I\equiv S_P=2eVG_0\sum_n\tau_n=2e<I>$, where $e$ is the electron charge, $G_0=2e^2/h$ is the conductance quantum, $\tau_n$ the transmission probability of the $n$\textsuperscript{th} channel, $I$ the current, and $h$ the Planck constant. The total conductance is defined by the sum over all transmission channels $G=G_0\sum_n\tau_n$. Here we assumed zero temperature ($T$), that the current follows Ohm's law ($I=VG$), and $\tau_n\ll1$ for all channels. 

This description of noise holds true when each electron passing through the junction finds an empty state at the other side. This is not the case for large junction transparencies when a forward-traveling electron finds an occupied state in the other lead and needs to be back-scattered for not violating Pauli's exclusion principle. This process effectively leads to a suppression of shot noise. The resulting noise can be described by the sum of products of transmissions and reflections: $S_I=2eVG_0\sum_n\tau_n(1-\tau_n)\le S_P$. 
The degree of shot-noise suppression is most conveniently expressed as the ratio between the measured noise and the Poissonian value. It is termed the Fano factor and defined as: $F\equiv S_I/S_P=\sum_n\tau_n(1-\tau_n)/\sum_n\tau_n$. It ranges between zero when all conduction channels are fully open, and 1 in the limit of $\tau_n\ll1$ as the noise approaches the Poissonian distribution. 
The Fano factor can be generalized for the transport of other charge carriers, carrying a charge of $q$. In this case, the Fano factor also stores information regarding the charge of the tunneling particle, $F\propto q/e$.

At finite temperatures, adopting a full quantum mechanical consideration results in a single equation for the current noise that includes both thermal- and shot-noise contributions:
\begin{equation}
	S_I=2G_0\big\{2k_BT\sum_{n}\tau_n^2+qV\coth(\frac{qV}{2k_BT})\sum_{n}\tau_n(1-\tau_n)\big\},
	\label{sn_eq}
\end{equation}
where $k_B$ is the Boltzmann constant, and the energy dependence of $\tau_n$ is neglected \cite{BLANTER2000}. At low voltages, $eV\ll k_BT$, Eq. \ref{sn_eq} is reduced to the well known thermal noise $S_{\text{Th}}=4k_BTG$. In contrast, the high-voltage limit is temperature independent, restoring the $T=0$ shot-noise dependence $S_I\approx2e<I>F$. This suggests that measuring at higher voltage values can separate shot noise from thermal noise. However, this solution is of limited use, because the system may exhibit a nonlinear current-voltage characteristic. Furthermore, since the $1/f$ noise increases quadratically with applied voltage, at higher voltages, it is likely to dominate the signal at larger frequencies.    

The first measurement of sub-Poissonian shot noise was reported from transport through quantum dots in 1995 by Reznikov \textit{et al.} \cite{Reznikov1995}. In the following years, shot noise was used to study various mesoscopic systems \cite{Picciotto1998,Saminadayar1997,Iannaccone1998,jehl2000,Cron2001,Oberholzer2001,Onac2006,Bid2009,Yamauchi2011,Altimiras2014,Ronen2015,ferrier2016}. Alongside mesoscopic measurements, molecules and atomic-scale structures have also been studied \cite{Brom1999,djukic2006,Kumar2012,Kumar2013,Karimi2016,Tewari2017,Tewari2019}, mainly by means of mechanically controlled break junctions. More recently a few experimental groups demonstrated local shot noise measurements using a scanning tunneling microscope (STM) \cite{Birk1995,kemiktarak2007,Herz2013,Burtzlaff2015,Bastiaans2018,Massee2018,Bastiaans2019,Koen2021,Mohr2019,Mohr2020,Mohr2021,thupakula2021}.

Similarly to measurements of mesoscopic systems, STM-based ones follow one of two main schemes: The first approach uses a room-temperature broadband amplifier connected in parallel to the current line to detect the shot noise signal \cite{Brom1999,Burtzlaff2015}. This technique is relatively easy to implement. The broadband noise amplifier is generally insensitive to the input resistance and allows to measure well above the $1/f$ noise. Its main drawback is that it requires post-measurement data correction to compensate for signal loss due to the low-pass filter formed by the tunnel junction resistance (\rj) and the relatively large capacitance at the input of the room-temperature amplifier related to the current wire's length. Furthermore, since many of the electronic components necessary for the measurement are situated in ambient conditions, thermal noise contributions are significant. The second approach uses a low-temperature resistance-inductance-capacitance (RLC) tank circuit and amplifier to sample the noise at the circuit's resonance frequency \cite{Reznikov1995,Bastiaans2018,Massee2018}. The circuit design is typically such that the resonance is in the MHz regime, well above the $1/f$ noise. This technique is very flexible in terms of the input resistance and can be used both in tunneling and contact (STM tip is touching the surface) modes. Moreover, due to the high frequencies used, averaging times are very short thus reducing the influence of mechanical instabilities and drifts. The drawback of this circuit design is that it requires a calibration of the amplification factor for each specific junction conductance value as \rj affects the amplification and resonance width. 

Here, we combine the advantages of amplification at low temperature and close to the junction using a broadband amplifier following developments in break-junction noise spectroscopy \cite{Tewari2017}. We insert such a circuit into a low-temperature STM, which allows us to measure single-atom junctions and to map the noise signal with atomic resolution. Our technique permits to measure at variable junction-resistance values, while profiting from a low thermal-noise background, and avoiding the need for data corrections due to the low input capacitance estimated to be 20 pF (see dashed black fit in Fig. \ref{RC}). Thus we are able to measure shot noise in tunnel junctions with conductance values ($G_J=1/R_J$) as low as 0.01 $G_0$ (see Fig. \ref{RC}). 

The article is structured as follows: we start by presenting our measurement setup in Sec. \ref{para_setup}. Next, we describe our measurement procedure in Sec. \ref{para_proc}, and demonstrate the apparatus capabilities by measuring shot noise on the (111) surface of gold. In Sec. \ref{para_disc} we finish with a short discussion of the results. 

\section{Shot-noise measurement setup} 
\label{para_setup}

\begin{figure*}[ht!]
	\includegraphics  [width=\linewidth] {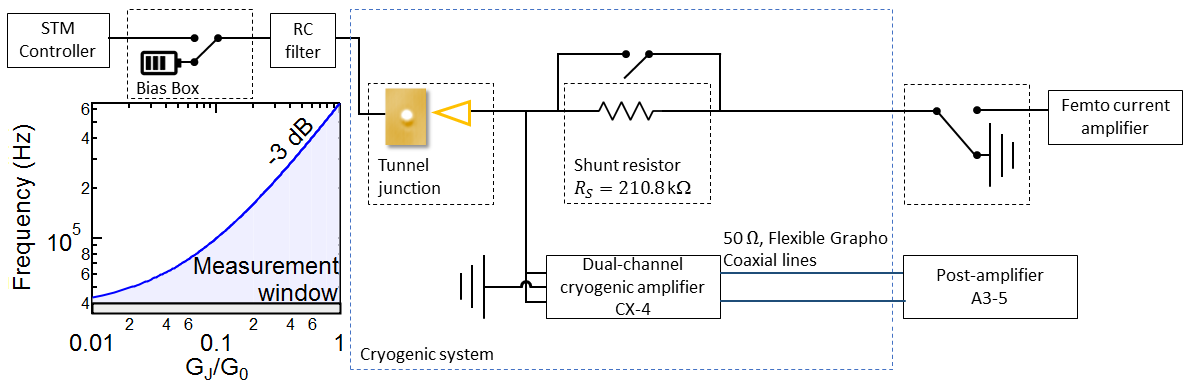}	
	\caption{\textbf{Block diagram of the shot-noise-measurement setup}. The STM junction and amplifier are mounted in UHV and are thermally connected to a liquid helium bath cryostat. Normal STM operation is controlled via a Nanonis SPM controller, which can be disconnected for noise measurements. Then, the voltage is supplied from batteries in a voltage bias box. The noise signal is detected between the tip-sample tunnel junction and a shunt resistor $R_S$, shunting the current drain line. The amplified noise is then transmitted via 50-$\Omega$ low-noise coaxial lines to a room-temperature post amplifier. To further reduce external noise during noise measurements, a remote-controlled grounding box is connected at the current output. In the bottom left we plot our relevant measurement window set by the low-temperature amplifier's frequency response, which is constant above the black line (see Fig. \ref{amp_freq}), and the -3 dB cutoff, blue line, related to the input circuit's effective RC filter. The blue line follows $f_{-3dB}=1/2\pi R_PC$, where $C=20$ pF. We can reliably measure shot noise within the shaded blue region between the two lines.   
	}
	\label{setup}
\end{figure*}

Our shot noise setup is integrated into a modified CreaTec STM operating under ultra-high vacuum (UHV) conditions and at a temperature of 4.3 K sustained by a liquid-helium bath cryostat. Normal scanning and spectroscopy modes of the STM remain fully operational in this integrated design. The main component of the modification is a low-temperature commercial broadband, dual-channel amplifier (Stahl electronics, model CX-4) installed close to the STM junction. (A block diagram of our setup is presented in Fig. \ref{setup}). The amplifier is mounted on a cold finger thermally shorted to the helium reservoir. A very good thermal contact is required between the amplifier and the cold finger in order to assure sufficient cooling power during operation. This is achieved by a thin indium foil. The dual-channel amplifier is connected via two parallel wires between the STM tip and a shunt resistor, $R_S$, shunting the current drain line. This shunting and the close proximity between tip and amplifier provide the required low input capacitance of the amplifier. An additional advantage of the setup geometry is that the close distance of the amplifier to the junction minimizes any pick-up of external noise prior to signal amplification. The amplifier effectively senses the voltage noise generated at the STM junction.

While the use of a shunt resistor is essential in our setup, it can potentially impair standard STM operation. It acts as a voltage divider for large tunnel-junction conductance values, and limits the current flow necessary for tip treatments (an input protection stage is designed in the low-temperature amplifier to protect it from rapid high voltage sweeps occurring during the tip treatments). Therefore we have installed a low temperature magnetic-latching RF relay (RF180) to bypass the shunt resistor. 

The two outputs of the low temperature amplifier are transmitted via 50-$\Omega$ flexible coaxial lines to a room-temperature, variable-gain, post-amplifier and cryo biasing unit (Stahl electronics, model A3-5). The insulator in these lines is graphite coated to reduce friction-related electrostatic noise. The two amplified signals are then recorded using a 14 bit National Instruments signal analyzer. A cross-correlation procedure is applied to remove uncorrelated noise picked-up in the two parallel signal lines. 

\begin{figure}[ht!]
	\includegraphics  [width=\linewidth] {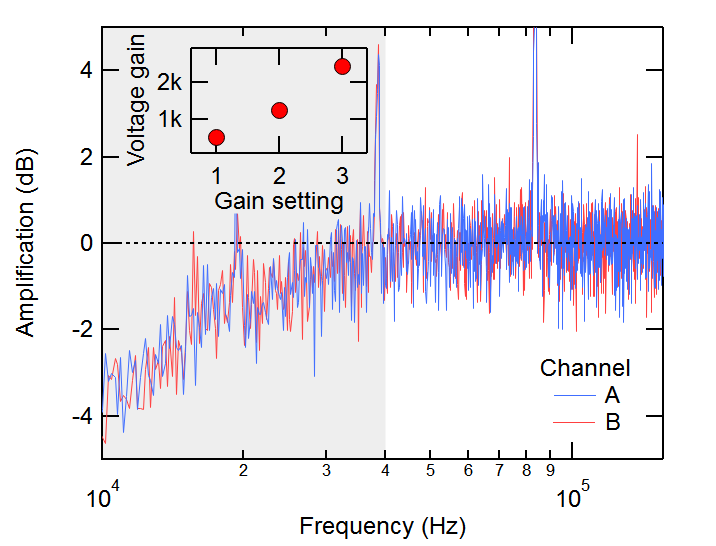}	
	\caption{\textbf{Noise amplification and frequency response}. Frequency-dependent amplification measured on the two channels of the low-temperature, broadband amplifier (labeled channel A (blue) and B (red)) as a function of a variable frequency supplied by a signal generator. Constant maximal gain is achieved above 40 kHz. Inset: Voltage gain of the noise amplification line vs. the different gain settings of the variable-gain post amplifier.
	}
	\label{amp_freq}
\end{figure}

The total amplification of the noise signal and its frequency dependence are measured \textit{in situ}. This is done by fully retracting the tip, grounding the voltage line, and introducing a small known AC excitation to the current drain line that is shorted to the low-temperature amplifier's input (the shunt resistor's bypass is closed to reduce damping during gain calibration). We measure the amplified signal (in units of $V/\sqrt{\text{Hz}}$) using our cross-correlation scheme at different gain settings of the post amplifier and divide the signal by the input amplitude to extract the overall gain (see inset of Fig. \ref{amp_freq}). We also sweep the input-signal frequency and measure the amplifier's response in both channels, see Fig. \ref{amp_freq}. The frequency response of both channels is in very good agreement, and is constant above $\sim40$ kHz.

The measured noise signal corresponds to a voltage noise at the STM junction. The corresponding current noise follows from the equivalent circuit diagram: $S_I=S_V/R_P^2$, where $R_P$ is the parallel resistance of the junction and shunt resistor $R_P=(1/R_s+dI/dV_j)^{-1}$, and $dI/dV_j$ is the differential conductance of the tunnel junction ($dI/dV_j\equiv G_J$ in the Ohmic regime). 
The determination of the noise signal thus requires knowledge of the applied voltage, the current, and the differential conductance ($dI/dV$). The current is measured using a room-temperature trans-impedance amplifier (Femto DLPCA-200), while the differential conductance is measured using a standard two-terminal lock-in technique. The noise conversion also requires precise knowledge of the shunt resistor. We therefore used a chip resistor that nominally varies from room temperature to liquid helium temperature by less than 5\%, and measured its low-temperature value by crashing the tip into the sample finding $R_S=210.8$ k$\Omega$. 
     
As the shot noise signal is usually very small the system has to exhibit extremely low external noise levels. Mechanical noise is reduced by standard STM noise-reduction schemes such as the use of pneumatic feet below the STM chamber and eddy-current damping of the STM head hanging on springs. Special care is further taken for filtering the electronic noise of all input signals. All cables connected to the STM are equipped with additional high-pass $\Pi$-filters, with the exception of the current and high-frequency lines. All UHV windows are covered by aluminum foil. Since the cleanest voltage signals are provided by batteries, we introduced a custom-built voltage-bias box that is remotely controlled and has a single input and two  electrically separated identical outputs. The two outputs are used for biasing the tunnel junction while independently monitoring the signal using the STM controller. The bias box can be switched between applying the output voltage from the STM controller for normal STM operation and a battery-powered low-noise voltage source that generates discrete values ranging between $\pm$300 mV used for noise measurements. A second remote-controlled grounding box is installed at the tip side (current drain), which grounds the current line during noise measurements. 

\section{Measurement procedure} 
\label{para_proc}
\begin{figure}[ht!]
	\includegraphics  [width=\linewidth] {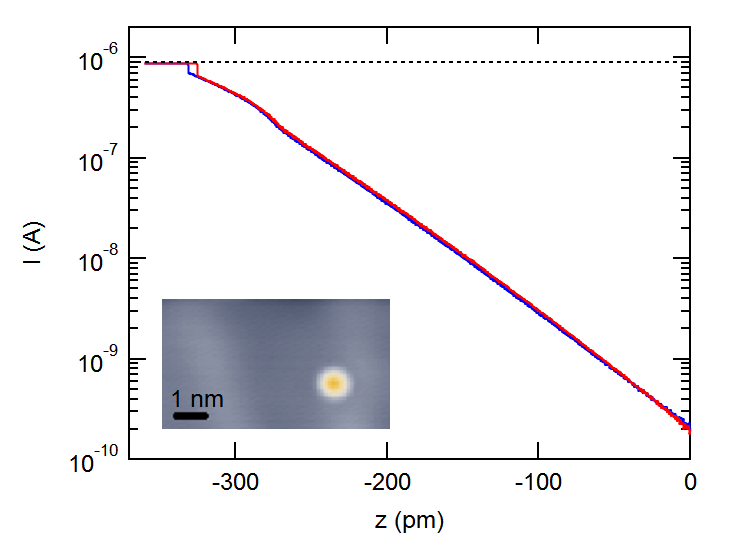}	
	\caption{\textbf{Approaching an adatom}. $I$, in log scale, vs $z$ curves for approaching (blue) and retracting (red) the tip to a gold adatom placed on the Au(111) surface. The initial exponential increase, linear in such log-linear plot, is terminated by a discontinuous "jump-to-contact", after which $I$ is nearly independent on $z$. Retracting the tip results in a hysteretic behavior for the current discontinuity, but an identical exponential dependence in the tunneling regime. The dashed black line represents the maximal possible current, which is limited by the shunt resistor $I_\text{max}=V/R_S$. Inset: Atom placed on the Au(111) surface, protruding 1.1 \AA\space from the Au(111) surface. The gray stripes are the result of the "herring-bone" reconstruction of the gold surface. A setpoint $I$=200 pA, $V$=200 mV is used for both the approach curves and the topography measurements.    
	}
	\label{approach}
\end{figure}

In this section we describe our measurement procedure and demonstrate our capabilities on a single gold atom placed on a Au(111) surface. Gold is chosen due to its well known electronic structure, having a single s-electron available for transport \cite{scheer1998}. We prepare the Au(111) surface using several sputtering-annealing cycles until a smooth surface exhibiting the well known "herring-bone" reconstruction is apparent \cite{Woll1989} (see inset instead of Fig. \ref{approach}).  

Successful shot-noise measurements require stringent stability of the STM tip-sample junction. As the shot noise amplitude directly depends on the current, an unstable junction resistance would lead to strongly fluctuating noise levels. The stability of an STM junction can be improved by adding an adatom on the surface. Contacting such adatom effectively minimizes the forces on the tip's apex compared to contacting a surface atom. 

Adatoms are created either by controllable tip indentation into the surface (400-500 pm using a set-point of $R_j=1$ G$\Omega$) or by adsorption of a dilute amount of adatoms from an external evaporator. A gold adatom deposited from a gold-coated tungsten tip is depicted in the inset of Fig. \ref{approach}. To check for stable and reproducible junction properties prior to noise measurements, we repetitively approach and retract the STM tip to/from the adatom and record the current during this procedure (see Fig \ref{approach}). Upon tip approach the tunneling current increases exponentially as expected for a tunnel junction. A sudden increase in the current signifies a "jump-to-contact". Further tip approach leaves the current nearly constant \cite{scheer1998}. If the junction is stable, the $I$-$z$ dependence during retraction is similar to that recorded during approach except for a small hysteresis in the transition from contact to tunnel regime. 

After testing the junction's stability, we set a desired junction resistance by approaching the STM tip to the corresponding $z$ position. Following a few minutes' delay to avoid drift, we commence the shot noise measurement cycle. As explained above, the Fano factor contains information of suppressed shot noise, allowing to deduce effects of correlations. To extract the Fano factor we measure the current dependence of the noise signal. This is done by recording the noise at a discrete set of equidistant voltage steps (positive and negative) supplied by the bias box. At each step we measure the current and voltage before grounding the current drain to measure the noise power ($2^{18}$ samples taken at a rate of 1 MS/s). We repeat the noise measurement and average the resulting power spectrum for at least 10 times (larger averaging reduces the standard deviation, but not the mean value of the signal). Thus, a single measurement cycle requires about 30 minutes. Once a measurement cycle is completed, the tip is retracted and a topographic image is recorded to confirm that the adatom's position and the tip's shape did not change during the measurement. This is done by cross-correlating the topographic images recorded before and after the shot noise acquisition routine. The measurement can then be repeated for different contact resistances (different $z$ position). To optimize the measurements, a LabVIEW program was developed to automatically execute the procedure described above.

\begin{figure}[ht!]
	\includegraphics  [width=\linewidth] {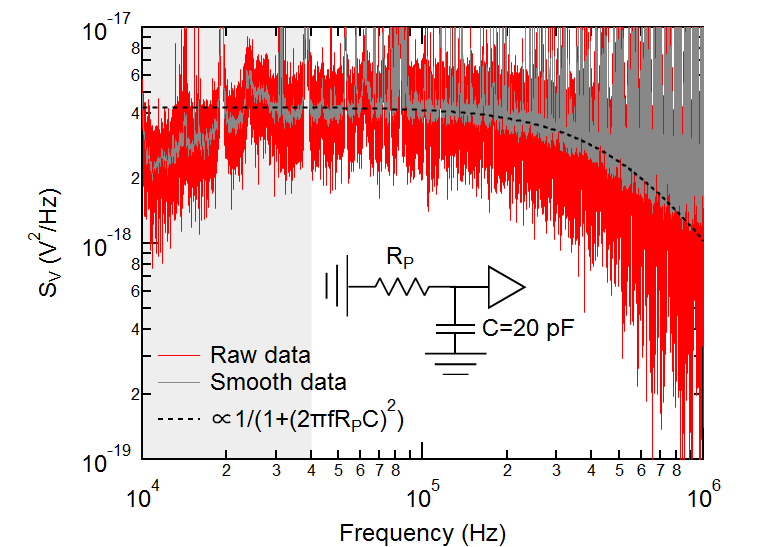}	
	\caption{\textbf{Noise data}. $S_V$ vs. frequency measured in contact conditions ($R_J=15.1~\text{k}\Omega$) at $V=0$. The data in red is the 10 time average of the two parallel noise measurements after cross-correlation. The gray curve is the same data after Gaussian smoothing, presented as a guide to the eye. The black dashed line is our best fit for the data above 40 kHz assuming only an RC cutoff of the input noise (see sketch). From the fit we can approximate the input capacitance of the low-temperature amplifier is to be $20$ pF. The shaded gray region indicates where the low-temperature amplifier's gain is not constant (see Fig. \ref{amp_freq}).
	}
	\label{RC}
\end{figure}

During data acquisition, the noise signal is recorded in the two parallel channels of the low-temperature amplifier, both shorted to the tip. The cross correlation of these two channels is computed directly in the LabVIEW program during the measurement's runtime. The full frequency-dependent cross-correlated $S_V$ acquired at $V=0$ at one particular junction is shown in Fig. \ref{RC}. Here, the tip was approached by 4 \AA\space starting from a set-point of $I=200$ pA and $V=200$ mV. Since shot noise is frequency independent, a constant noise plateau should be observable whenever the gain of the low temperature amplifier is constant. This is true above 40 kHz (in agreement with Fig. \ref{amp_freq}), and below $\sim$500 kHz. The high frequency cutoff is related to the -3 dB point of the low-pass filter formed by the junction's resistance and unavoidable input capacitance. It is determined by fitting $S_V$ to the expected frequency response of a low-pass RC filter, $S_V\propto(1+(2\pi fR_PC)^2)^{-1}$ (see sketch within Fig. \ref{RC}). We find, as previously mentioned, $C\approx20$ pF. In the bottom left of Fig. \ref{setup} we use this capacitance value to sketch our measurement window as a function of junction conductance. The blue line in the figure follows the -3 dB point of an RC filter, $f_{-3dB}=1/2\pi R_PC$, and the black line the onset of full amplification. Yet, since the signal loss at the -3 dB point is already significant, we average the noise signal at much lower frequencies (usually between 90-100 kHz). Within such junction-dependent frequency window, $S_V$ is evaluated with less than 3\% deviation. Note that the 1/f noise contribution is indeed suppressed within our measurement window. We further make sure to avoid the influence of the narrow peaks observed in the noise curves. These noise peaks normally stem from power supplies and are created by their switching-mode operation. 

\begin{figure*}[ht!]
	\includegraphics  [width=17 cm] {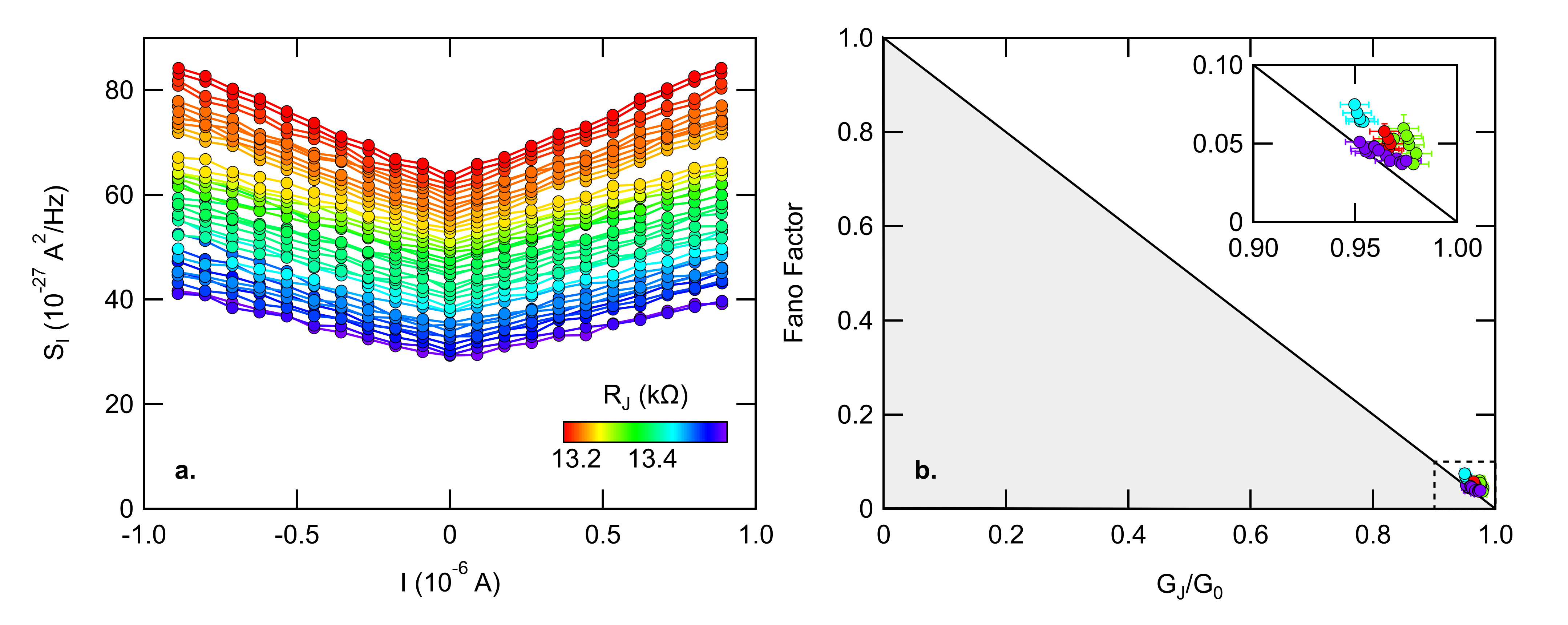}	
	\caption{\textbf{Current noise and Fano factor}. \textbf{a}.: $S_I$ vs $I$ curves, shifted vertically for visibility. The different colors represent the junctions' resistance values. The data follows the expected linear dependence in $I$. \textbf{b}.: Fano factor vs the normalized junction conductance. The black line follows the theoretical prediction for a single spin-degenerate transport channel as is the case of gold. Data should only be above the black line. Inset: zoom in on the data.    
	}
	\label{fano}
\end{figure*}

As explained above, the Fano factor follows $S_I=2e<I>F$ in the high-voltage limit, which is valid whenever $k_BT\ll eV$. Hence, shot noise is expected to increase linearly with the DC current with a slope determined by the Fano factor. In Fig. \ref{fano}\textbf{a}. we plot $S_I$ vs $I$ measured on different single gold atom junctions. The curves, shifted vertically for visibility, are colored according to the junctions' resistance values and follow the expected linear trend. 
We note that, while the thermal noise measured at $V=0$ is linearly dependent on $1/R_P$ we observe a small offset: $S_I-S_{\text{Th}}\approx 6.1\times10^{-27}$ A\textsuperscript{2}/Hz, evaluated at $R_P=13.5$ k$\Omega$. This offset cannot be explained by the small input voltage noise of the low temperature amplifier (nominally $v_n\approx0.6\text{ nV} /\sqrt{\text{Hz}}$). We may speculate that the temperature at the junction is slightly higher than the one measured (at the base plate of the STM). A temperature increase of 1.5 K would be in very good agreement with the measured offset, but might not necessarily be its origin. In any case, this offset does not affect the determination of the Fano factor as the latter is given by the slope, which is temperature independent.

For each curve we evaluate the Fano factor for positive and negative currents and plot the averaged value in Fig. \ref{fano}\textbf{b}. The error in evaluating the Fano factor is estimated to be half the difference between the two independent fit values. The error in the normalized conductance ($G_J/G_0$) is estimated as $\delta G_J/G=100 \Omega/ R_J$, with 100 $\Omega$ being our precision in measuring the resistance. The black line in the figure corresponds to the theoretical dependence of the Fano factor in the case of a single channel spin-degenerate transport, as expected for single-gold-atom junctions \cite{scheer1998,BLANTER2000}. We find very good agreement between experiment and theory. This shows that our setup is fully operational for shot noise measurements. 

\section{Discussion}
\label{para_disc}
We presented a measurement scheme enabling the local measurements of shot noise integrated in a fully functional low-temperature UHV STM apparatus. We have demonstrated our capabilities measuring the Fano factor of single-atom junctions by contacting gold adatoms on a Au(111) surface. The great advantage of our setup is its flexibility in \rj that will allow future measurements in non-linear junctions, \textit{e.g.} superconducting junctions. 
More generally, since shot noise is a direct probe for the charge of the tunneling quasi-particle, such measurement technique holds great promise in providing "smoking gun" evidence of the existence of exotic quasi-particle states. 

Finally, while shot noise seems to be a valuable observable for the study of fundamental physical processes, it may also be considered nuisance for nanoscale electronic devices.  Its independence of temperature and frequency makes it the main source of noise in devices operating at low temperatures. Hence, besides the fundamental insights that may be gained from shot noise measurements, its characteristics in nanoscopic junctions has also technological implications.

\section*{Acknowledgments}
We would like to thank Jan van Ruitenbeek and Manohar Kumar for discussions at the very beginning of the project, Stefan Stahl who designed the low-temperature amplifier and contributed valuable input during the setup stages of the system.  
I.T. acknowledges funding from the Alexander-von-Humboldt
foundation in the framework of Humboldt Research Fellowship
for Postdoctoral Researchers, and from the DFG in the framework of the Walter Benjamin Position (TA 1722/1-1). We also acknowledge funding by the European Research Council through the Consolidator Grant "NanoSpin" (project number 616623) (K.J.F.). 

\section*{Data Availability}
The data that support the findings of this study are available from the corresponding author upon reasonable request.

\section*{Competing financial interests}
The authors declare no competing financial interests.

\bibliographystyle{apsrev4-2}

\bibliography{snpaper}

\begin{thebibliography}{44}%
\makeatletter
\providecommand \@ifxundefined [1]{%
 \@ifx{#1\undefined}
}%
\providecommand \@ifnum [1]{%
 \ifnum #1\expandafter \@firstoftwo
 \else \expandafter \@secondoftwo
 \fi
}%
\providecommand \@ifx [1]{%
 \ifx #1\expandafter \@firstoftwo
 \else \expandafter \@secondoftwo
 \fi
}%
\providecommand \natexlab [1]{#1}%
\providecommand \enquote  [1]{``#1''}%
\providecommand \bibnamefont  [1]{#1}%
\providecommand \bibfnamefont [1]{#1}%
\providecommand \citenamefont [1]{#1}%
\providecommand \href@noop [0]{\@secondoftwo}%
\providecommand \href [0]{\begingroup \@sanitize@url \@href}%
\providecommand \@href[1]{\@@startlink{#1}\@@href}%
\providecommand \@@href[1]{\endgroup#1\@@endlink}%
\providecommand \@sanitize@url [0]{\catcode `\\12\catcode `\$12\catcode
  `\&12\catcode `\#12\catcode `\^12\catcode `\_12\catcode `\%12\relax}%
\providecommand \@@startlink[1]{}%
\providecommand \@@endlink[0]{}%
\providecommand \url  [0]{\begingroup\@sanitize@url \@url }%
\providecommand \@url [1]{\endgroup\@href {#1}{\urlprefix }}%
\providecommand \urlprefix  [0]{URL }%
\providecommand \Eprint [0]{\href }%
\providecommand \doibase [0]{https://doi.org/}%
\providecommand \selectlanguage [0]{\@gobble}%
\providecommand \bibinfo  [0]{\@secondoftwo}%
\providecommand \bibfield  [0]{\@secondoftwo}%
\providecommand \translation [1]{[#1]}%
\providecommand \BibitemOpen [0]{}%
\providecommand \bibitemStop [0]{}%
\providecommand \bibitemNoStop [0]{.\EOS\space}%
\providecommand \EOS [0]{\spacefactor3000\relax}%
\providecommand \BibitemShut  [1]{\csname bibitem#1\endcsname}%
\let\auto@bib@innerbib\@empty
%</preamble>
\bibitem [{\citenamefont {Schottky}(2018)}]{Schottky}%
  \BibitemOpen
  \bibfield  {author} {\bibinfo {author} {\bibfnamefont {W.}~\bibnamefont
  {Schottky}},\ }\href {https://doi.org/10.1117/1.JMM.17.4.041001} {\bibfield
  {journal} {\bibinfo  {journal} {Journal of Micro/Nanolithography, MEMS, and
  MOEMS}\ }\textbf {\bibinfo {volume} {17}},\ \bibinfo {pages} {1 } (\bibinfo
  {year} {2018})}\BibitemShut {NoStop}%
\bibitem [{\citenamefont {Blanter}\ and\ \citenamefont
  {Büttiker}(2000)}]{BLANTER2000}%
  \BibitemOpen
  \bibfield  {author} {\bibinfo {author} {\bibfnamefont {Y.}~\bibnamefont
  {Blanter}}\ and\ \bibinfo {author} {\bibfnamefont {M.}~\bibnamefont
  {Büttiker}},\ }\href
  {https://doi.org/https://doi.org/10.1016/S0370-1573(99)00123-4} {\bibfield
  {journal} {\bibinfo  {journal} {Physics Reports}\ }\textbf {\bibinfo {volume}
  {336}},\ \bibinfo {pages} {1} (\bibinfo {year} {2000})}\BibitemShut {NoStop}%
\bibitem [{\citenamefont {Saminadayar}\ \emph {et~al.}(1997)\citenamefont
  {Saminadayar}, \citenamefont {Glattli}, \citenamefont {Jin},\ and\
  \citenamefont {Etienne}}]{Saminadayar1997}%
  \BibitemOpen
  \bibfield  {author} {\bibinfo {author} {\bibfnamefont {L.}~\bibnamefont
  {Saminadayar}}, \bibinfo {author} {\bibfnamefont {D.~C.}\ \bibnamefont
  {Glattli}}, \bibinfo {author} {\bibfnamefont {Y.}~\bibnamefont {Jin}},\ and\
  \bibinfo {author} {\bibfnamefont {B.}~\bibnamefont {Etienne}},\ }\href
  {https://doi.org/10.1103/PhysRevLett.79.2526} {\bibfield  {journal} {\bibinfo
   {journal} {Phys. Rev. Lett.}\ }\textbf {\bibinfo {volume} {79}},\ \bibinfo
  {pages} {2526} (\bibinfo {year} {1997})}\BibitemShut {NoStop}%
\bibitem [{\citenamefont {De-Picciotto}\ \emph {et~al.}(1998)\citenamefont
  {De-Picciotto}, \citenamefont {Reznikov}, \citenamefont {Heiblum},
  \citenamefont {Umansky}, \citenamefont {Bunin},\ and\ \citenamefont
  {Mahalu}}]{Picciotto1998}%
  \BibitemOpen
  \bibfield  {author} {\bibinfo {author} {\bibfnamefont {R.}~\bibnamefont
  {De-Picciotto}}, \bibinfo {author} {\bibfnamefont {M.}~\bibnamefont
  {Reznikov}}, \bibinfo {author} {\bibfnamefont {M.}~\bibnamefont {Heiblum}},
  \bibinfo {author} {\bibfnamefont {V.}~\bibnamefont {Umansky}}, \bibinfo
  {author} {\bibfnamefont {G.}~\bibnamefont {Bunin}},\ and\ \bibinfo {author}
  {\bibfnamefont {D.}~\bibnamefont {Mahalu}},\ }\href
  {https://doi.org/https://doi.org/10.1038/38241} {\bibfield  {journal}
  {\bibinfo  {journal} {Physica B: Condensed Matter}\ }\textbf {\bibinfo
  {volume} {249}},\ \bibinfo {pages} {395} (\bibinfo {year}
  {1998})}\BibitemShut {NoStop}%
\bibitem [{\citenamefont {Bid}\ \emph {et~al.}(2009)\citenamefont {Bid},
  \citenamefont {Ofek}, \citenamefont {Heiblum}, \citenamefont {Umansky},\ and\
  \citenamefont {Mahalu}}]{Bid2009}%
  \BibitemOpen
  \bibfield  {author} {\bibinfo {author} {\bibfnamefont {A.}~\bibnamefont
  {Bid}}, \bibinfo {author} {\bibfnamefont {N.}~\bibnamefont {Ofek}}, \bibinfo
  {author} {\bibfnamefont {M.}~\bibnamefont {Heiblum}}, \bibinfo {author}
  {\bibfnamefont {V.}~\bibnamefont {Umansky}},\ and\ \bibinfo {author}
  {\bibfnamefont {D.}~\bibnamefont {Mahalu}},\ }\href
  {https://doi.org/10.1103/PhysRevLett.103.236802} {\bibfield  {journal}
  {\bibinfo  {journal} {Phys. Rev. Lett.}\ }\textbf {\bibinfo {volume} {103}},\
  \bibinfo {pages} {236802} (\bibinfo {year} {2009})}\BibitemShut {NoStop}%
\bibitem [{\citenamefont {Jehl}\ \emph {et~al.}(2000)\citenamefont {Jehl},
  \citenamefont {Sanquer}, \citenamefont {Calemczuk},\ and\ \citenamefont
  {Mailly}}]{jehl2000}%
  \BibitemOpen
  \bibfield  {author} {\bibinfo {author} {\bibfnamefont {X.}~\bibnamefont
  {Jehl}}, \bibinfo {author} {\bibfnamefont {M.}~\bibnamefont {Sanquer}},
  \bibinfo {author} {\bibfnamefont {R.}~\bibnamefont {Calemczuk}},\ and\
  \bibinfo {author} {\bibfnamefont {D.}~\bibnamefont {Mailly}},\ }\href
  {https://doi.org/https://doi.org/10.1038/35011012} {\bibfield  {journal}
  {\bibinfo  {journal} {Nature}\ }\textbf {\bibinfo {volume} {405}},\ \bibinfo
  {pages} {50} (\bibinfo {year} {2000})}\BibitemShut {NoStop}%
\bibitem [{\citenamefont {Bastiaans}\ \emph {et~al.}(2019)\citenamefont
  {Bastiaans}, \citenamefont {Cho}, \citenamefont {Chatzopoulos}, \citenamefont
  {Leeuwenhoek}, \citenamefont {Koks},\ and\ \citenamefont
  {Allan}}]{Bastiaans2019}%
  \BibitemOpen
  \bibfield  {author} {\bibinfo {author} {\bibfnamefont {K.~M.}\ \bibnamefont
  {Bastiaans}}, \bibinfo {author} {\bibfnamefont {D.}~\bibnamefont {Cho}},
  \bibinfo {author} {\bibfnamefont {D.}~\bibnamefont {Chatzopoulos}}, \bibinfo
  {author} {\bibfnamefont {M.}~\bibnamefont {Leeuwenhoek}}, \bibinfo {author}
  {\bibfnamefont {C.}~\bibnamefont {Koks}},\ and\ \bibinfo {author}
  {\bibfnamefont {M.~P.}\ \bibnamefont {Allan}},\ }\href
  {https://doi.org/10.1103/PhysRevB.100.104506} {\bibfield  {journal} {\bibinfo
   {journal} {Phys. Rev. B}\ }\textbf {\bibinfo {volume} {100}},\ \bibinfo
  {pages} {104506} (\bibinfo {year} {2019})}\BibitemShut {NoStop}%
\bibitem [{\citenamefont {Bastiaans}\ \emph {et~al.}(2021)\citenamefont
  {Bastiaans}, \citenamefont {Chatzopoulos}, \citenamefont {Ge}, \citenamefont
  {Cho}, \citenamefont {Tromp}, \citenamefont {van Ruitenbeek}, \citenamefont
  {Fischer}, \citenamefont {de~Visser}, \citenamefont {Thoen}, \citenamefont
  {Driessen}, \citenamefont {Klapwijk},\ and\ \citenamefont
  {Allan}}]{Koen2021}%
  \BibitemOpen
  \bibfield  {author} {\bibinfo {author} {\bibfnamefont {K.~M.}\ \bibnamefont
  {Bastiaans}}, \bibinfo {author} {\bibfnamefont {D.}~\bibnamefont
  {Chatzopoulos}}, \bibinfo {author} {\bibfnamefont {J.-F.}\ \bibnamefont
  {Ge}}, \bibinfo {author} {\bibfnamefont {D.}~\bibnamefont {Cho}}, \bibinfo
  {author} {\bibfnamefont {W.~O.}\ \bibnamefont {Tromp}}, \bibinfo {author}
  {\bibfnamefont {J.~M.}\ \bibnamefont {van Ruitenbeek}}, \bibinfo {author}
  {\bibfnamefont {M.~H.}\ \bibnamefont {Fischer}}, \bibinfo {author}
  {\bibfnamefont {P.~J.}\ \bibnamefont {de~Visser}}, \bibinfo {author}
  {\bibfnamefont {D.~J.}\ \bibnamefont {Thoen}}, \bibinfo {author}
  {\bibfnamefont {E.~F.~C.}\ \bibnamefont {Driessen}}, \bibinfo {author}
  {\bibfnamefont {T.~M.}\ \bibnamefont {Klapwijk}},\ and\ \bibinfo {author}
  {\bibfnamefont {M.~P.}\ \bibnamefont {Allan}},\ }\href
  {https://doi.org/10.1126/science.abe3987} {\bibfield  {journal} {\bibinfo
  {journal} {Science}\ }\textbf {\bibinfo {volume} {374}},\ \bibinfo {pages}
  {608} (\bibinfo {year} {2021})}\BibitemShut {NoStop}%
\bibitem [{\citenamefont {Meir}\ and\ \citenamefont {Golub}(2002)}]{Meir2002}%
  \BibitemOpen
  \bibfield  {author} {\bibinfo {author} {\bibfnamefont {Y.}~\bibnamefont
  {Meir}}\ and\ \bibinfo {author} {\bibfnamefont {A.}~\bibnamefont {Golub}},\
  }\href {https://doi.org/10.1103/PhysRevLett.88.116802} {\bibfield  {journal}
  {\bibinfo  {journal} {Phys. Rev. Lett.}\ }\textbf {\bibinfo {volume} {88}},\
  \bibinfo {pages} {116802} (\bibinfo {year} {2002})}\BibitemShut {NoStop}%
\bibitem [{\citenamefont {Sela}\ \emph {et~al.}(2006)\citenamefont {Sela},
  \citenamefont {Oreg}, \citenamefont {von Oppen},\ and\ \citenamefont
  {Koch}}]{Sela2006}%
  \BibitemOpen
  \bibfield  {author} {\bibinfo {author} {\bibfnamefont {E.}~\bibnamefont
  {Sela}}, \bibinfo {author} {\bibfnamefont {Y.}~\bibnamefont {Oreg}}, \bibinfo
  {author} {\bibfnamefont {F.}~\bibnamefont {von Oppen}},\ and\ \bibinfo
  {author} {\bibfnamefont {J.}~\bibnamefont {Koch}},\ }\href
  {https://doi.org/10.1103/PhysRevLett.97.086601} {\bibfield  {journal}
  {\bibinfo  {journal} {Phys. Rev. Lett.}\ }\textbf {\bibinfo {volume} {97}},\
  \bibinfo {pages} {086601} (\bibinfo {year} {2006})}\BibitemShut {NoStop}%
\bibitem [{\citenamefont {Zarchin}\ \emph {et~al.}(2008)\citenamefont
  {Zarchin}, \citenamefont {Zaffalon}, \citenamefont {Heiblum}, \citenamefont
  {Mahalu},\ and\ \citenamefont {Umansky}}]{Zarchin2008}%
  \BibitemOpen
  \bibfield  {author} {\bibinfo {author} {\bibfnamefont {O.}~\bibnamefont
  {Zarchin}}, \bibinfo {author} {\bibfnamefont {M.}~\bibnamefont {Zaffalon}},
  \bibinfo {author} {\bibfnamefont {M.}~\bibnamefont {Heiblum}}, \bibinfo
  {author} {\bibfnamefont {D.}~\bibnamefont {Mahalu}},\ and\ \bibinfo {author}
  {\bibfnamefont {V.}~\bibnamefont {Umansky}},\ }\href
  {https://doi.org/10.1103/PhysRevB.77.241303} {\bibfield  {journal} {\bibinfo
  {journal} {Phys. Rev. B}\ }\textbf {\bibinfo {volume} {77}},\ \bibinfo
  {pages} {241303} (\bibinfo {year} {2008})}\BibitemShut {NoStop}%
\bibitem [{\citenamefont {Delattre}\ \emph {et~al.}(2009)\citenamefont
  {Delattre}, \citenamefont {Feuillet-Palma}, \citenamefont {Herrmann},
  \citenamefont {Morfin}, \citenamefont {Berroir}, \citenamefont {F{\`e}ve},
  \citenamefont {Pla{\c{c}}ais}, \citenamefont {Glattli}, \citenamefont {Choi},
  \citenamefont {Mora} \emph {et~al.}}]{delattre2009}%
  \BibitemOpen
  \bibfield  {author} {\bibinfo {author} {\bibfnamefont {T.}~\bibnamefont
  {Delattre}}, \bibinfo {author} {\bibfnamefont {C.}~\bibnamefont
  {Feuillet-Palma}}, \bibinfo {author} {\bibfnamefont {L.}~\bibnamefont
  {Herrmann}}, \bibinfo {author} {\bibfnamefont {P.}~\bibnamefont {Morfin}},
  \bibinfo {author} {\bibfnamefont {J.-M.}\ \bibnamefont {Berroir}}, \bibinfo
  {author} {\bibfnamefont {G.}~\bibnamefont {F{\`e}ve}}, \bibinfo {author}
  {\bibfnamefont {B.}~\bibnamefont {Pla{\c{c}}ais}}, \bibinfo {author}
  {\bibfnamefont {D.}~\bibnamefont {Glattli}}, \bibinfo {author} {\bibfnamefont
  {M.-S.}\ \bibnamefont {Choi}}, \bibinfo {author} {\bibfnamefont
  {C.}~\bibnamefont {Mora}}, \emph {et~al.},\ }\href
  {https://doi.org/https://doi.org/10.1038/nphys1186} {\bibfield  {journal}
  {\bibinfo  {journal} {Nature Physics}\ }\textbf {\bibinfo {volume} {5}},\
  \bibinfo {pages} {208} (\bibinfo {year} {2009})}\BibitemShut {NoStop}%
\bibitem [{\citenamefont {Yamauchi}\ \emph {et~al.}(2011)\citenamefont
  {Yamauchi}, \citenamefont {Sekiguchi}, \citenamefont {Chida}, \citenamefont
  {Arakawa}, \citenamefont {Nakamura}, \citenamefont {Kobayashi}, \citenamefont
  {Ono}, \citenamefont {Fujii},\ and\ \citenamefont {Sakano}}]{Yamauchi2011}%
  \BibitemOpen
  \bibfield  {author} {\bibinfo {author} {\bibfnamefont {Y.}~\bibnamefont
  {Yamauchi}}, \bibinfo {author} {\bibfnamefont {K.}~\bibnamefont {Sekiguchi}},
  \bibinfo {author} {\bibfnamefont {K.}~\bibnamefont {Chida}}, \bibinfo
  {author} {\bibfnamefont {T.}~\bibnamefont {Arakawa}}, \bibinfo {author}
  {\bibfnamefont {S.}~\bibnamefont {Nakamura}}, \bibinfo {author}
  {\bibfnamefont {K.}~\bibnamefont {Kobayashi}}, \bibinfo {author}
  {\bibfnamefont {T.}~\bibnamefont {Ono}}, \bibinfo {author} {\bibfnamefont
  {T.}~\bibnamefont {Fujii}},\ and\ \bibinfo {author} {\bibfnamefont
  {R.}~\bibnamefont {Sakano}},\ }\href
  {https://doi.org/10.1103/PhysRevLett.106.176601} {\bibfield  {journal}
  {\bibinfo  {journal} {Phys. Rev. Lett.}\ }\textbf {\bibinfo {volume} {106}},\
  \bibinfo {pages} {176601} (\bibinfo {year} {2011})}\BibitemShut {NoStop}%
\bibitem [{\citenamefont {Burtzlaff}\ \emph {et~al.}(2015)\citenamefont
  {Burtzlaff}, \citenamefont {Weismann}, \citenamefont {Brandbyge},\ and\
  \citenamefont {Berndt}}]{Burtzlaff2015}%
  \BibitemOpen
  \bibfield  {author} {\bibinfo {author} {\bibfnamefont {A.}~\bibnamefont
  {Burtzlaff}}, \bibinfo {author} {\bibfnamefont {A.}~\bibnamefont {Weismann}},
  \bibinfo {author} {\bibfnamefont {M.}~\bibnamefont {Brandbyge}},\ and\
  \bibinfo {author} {\bibfnamefont {R.}~\bibnamefont {Berndt}},\ }\href
  {https://doi.org/10.1103/PhysRevLett.114.016602} {\bibfield  {journal}
  {\bibinfo  {journal} {Phys. Rev. Lett.}\ }\textbf {\bibinfo {volume} {114}},\
  \bibinfo {pages} {016602} (\bibinfo {year} {2015})}\BibitemShut {NoStop}%
\bibitem [{\citenamefont {Ferrier}\ \emph {et~al.}(2016)\citenamefont
  {Ferrier}, \citenamefont {Arakawa}, \citenamefont {Hata}, \citenamefont
  {Fujiwara}, \citenamefont {Delagrange}, \citenamefont {Weil}, \citenamefont
  {Deblock}, \citenamefont {Sakano}, \citenamefont {Oguri},\ and\ \citenamefont
  {Kobayashi}}]{ferrier2016}%
  \BibitemOpen
  \bibfield  {author} {\bibinfo {author} {\bibfnamefont {M.}~\bibnamefont
  {Ferrier}}, \bibinfo {author} {\bibfnamefont {T.}~\bibnamefont {Arakawa}},
  \bibinfo {author} {\bibfnamefont {T.}~\bibnamefont {Hata}}, \bibinfo {author}
  {\bibfnamefont {R.}~\bibnamefont {Fujiwara}}, \bibinfo {author}
  {\bibfnamefont {R.}~\bibnamefont {Delagrange}}, \bibinfo {author}
  {\bibfnamefont {R.}~\bibnamefont {Weil}}, \bibinfo {author} {\bibfnamefont
  {R.}~\bibnamefont {Deblock}}, \bibinfo {author} {\bibfnamefont
  {R.}~\bibnamefont {Sakano}}, \bibinfo {author} {\bibfnamefont
  {A.}~\bibnamefont {Oguri}},\ and\ \bibinfo {author} {\bibfnamefont
  {K.}~\bibnamefont {Kobayashi}},\ }\href
  {https://doi.org/https://doi.org/10.1038/nphys3556} {\bibfield  {journal}
  {\bibinfo  {journal} {Nature Physics}\ }\textbf {\bibinfo {volume} {12}},\
  \bibinfo {pages} {230} (\bibinfo {year} {2016})}\BibitemShut {NoStop}%
\bibitem [{\citenamefont {Cocklin}\ and\ \citenamefont
  {Morr}(2019)}]{Cocklin2019}%
  \BibitemOpen
  \bibfield  {author} {\bibinfo {author} {\bibfnamefont {S.}~\bibnamefont
  {Cocklin}}\ and\ \bibinfo {author} {\bibfnamefont {D.~K.}\ \bibnamefont
  {Morr}},\ }\href {https://doi.org/10.1103/PhysRevB.100.125146} {\bibfield
  {journal} {\bibinfo  {journal} {Phys. Rev. B}\ }\textbf {\bibinfo {volume}
  {100}},\ \bibinfo {pages} {125146} (\bibinfo {year} {2019})}\BibitemShut
  {NoStop}%
\bibitem [{\citenamefont {Mohr}\ \emph {et~al.}(2020)\citenamefont {Mohr},
  \citenamefont {Gruber}, \citenamefont {Weismann}, \citenamefont {Jacob},
  \citenamefont {Abufager}, \citenamefont {Lorente},\ and\ \citenamefont
  {Berndt}}]{Mohr2020}%
  \BibitemOpen
  \bibfield  {author} {\bibinfo {author} {\bibfnamefont {M.}~\bibnamefont
  {Mohr}}, \bibinfo {author} {\bibfnamefont {M.}~\bibnamefont {Gruber}},
  \bibinfo {author} {\bibfnamefont {A.}~\bibnamefont {Weismann}}, \bibinfo
  {author} {\bibfnamefont {D.}~\bibnamefont {Jacob}}, \bibinfo {author}
  {\bibfnamefont {P.}~\bibnamefont {Abufager}}, \bibinfo {author}
  {\bibfnamefont {N.}~\bibnamefont {Lorente}},\ and\ \bibinfo {author}
  {\bibfnamefont {R.}~\bibnamefont {Berndt}},\ }\href
  {https://doi.org/10.1103/PhysRevB.101.075414} {\bibfield  {journal} {\bibinfo
   {journal} {Phys. Rev. B}\ }\textbf {\bibinfo {volume} {101}},\ \bibinfo
  {pages} {075414} (\bibinfo {year} {2020})}\BibitemShut {NoStop}%
\bibitem [{\citenamefont {Johnson}(1928)}]{Johnson1928}%
  \BibitemOpen
  \bibfield  {author} {\bibinfo {author} {\bibfnamefont {J.~B.}\ \bibnamefont
  {Johnson}},\ }\href {https://doi.org/10.1103/PhysRev.32.97} {\bibfield
  {journal} {\bibinfo  {journal} {Phys. Rev.}\ }\textbf {\bibinfo {volume}
  {32}},\ \bibinfo {pages} {97} (\bibinfo {year} {1928})}\BibitemShut {NoStop}%
\bibitem [{\citenamefont {Nyquist}(1928)}]{Nyquist1928}%
  \BibitemOpen
  \bibfield  {author} {\bibinfo {author} {\bibfnamefont {H.}~\bibnamefont
  {Nyquist}},\ }\href {https://doi.org/10.1103/PhysRev.32.110} {\bibfield
  {journal} {\bibinfo  {journal} {Phys. Rev.}\ }\textbf {\bibinfo {volume}
  {32}},\ \bibinfo {pages} {110} (\bibinfo {year} {1928})}\BibitemShut
  {NoStop}%
\bibitem [{\citenamefont {Johnson}(1925)}]{Johnson1925}%
  \BibitemOpen
  \bibfield  {author} {\bibinfo {author} {\bibfnamefont {J.~B.}\ \bibnamefont
  {Johnson}},\ }\href {https://doi.org/10.1103/PhysRev.26.71} {\bibfield
  {journal} {\bibinfo  {journal} {Phys. Rev.}\ }\textbf {\bibinfo {volume}
  {26}},\ \bibinfo {pages} {71} (\bibinfo {year} {1925})}\BibitemShut {NoStop}%
\bibitem [{\citenamefont {Reznikov}\ \emph {et~al.}(1995)\citenamefont
  {Reznikov}, \citenamefont {Heiblum}, \citenamefont {Shtrikman},\ and\
  \citenamefont {Mahalu}}]{Reznikov1995}%
  \BibitemOpen
  \bibfield  {author} {\bibinfo {author} {\bibfnamefont {M.}~\bibnamefont
  {Reznikov}}, \bibinfo {author} {\bibfnamefont {M.}~\bibnamefont {Heiblum}},
  \bibinfo {author} {\bibfnamefont {H.}~\bibnamefont {Shtrikman}},\ and\
  \bibinfo {author} {\bibfnamefont {D.}~\bibnamefont {Mahalu}},\ }\href
  {https://doi.org/10.1103/PhysRevLett.75.3340} {\bibfield  {journal} {\bibinfo
   {journal} {Phys. Rev. Lett.}\ }\textbf {\bibinfo {volume} {75}},\ \bibinfo
  {pages} {3340} (\bibinfo {year} {1995})}\BibitemShut {NoStop}%
\bibitem [{\citenamefont {Iannaccone}\ \emph {et~al.}(1998)\citenamefont
  {Iannaccone}, \citenamefont {Lombardi}, \citenamefont {Macucci},\ and\
  \citenamefont {Pellegrini}}]{Iannaccone1998}%
  \BibitemOpen
  \bibfield  {author} {\bibinfo {author} {\bibfnamefont {G.}~\bibnamefont
  {Iannaccone}}, \bibinfo {author} {\bibfnamefont {G.}~\bibnamefont
  {Lombardi}}, \bibinfo {author} {\bibfnamefont {M.}~\bibnamefont {Macucci}},\
  and\ \bibinfo {author} {\bibfnamefont {B.}~\bibnamefont {Pellegrini}},\
  }\href {https://doi.org/10.1103/PhysRevLett.80.1054} {\bibfield  {journal}
  {\bibinfo  {journal} {Phys. Rev. Lett.}\ }\textbf {\bibinfo {volume} {80}},\
  \bibinfo {pages} {1054} (\bibinfo {year} {1998})}\BibitemShut {NoStop}%
\bibitem [{\citenamefont {Cron}\ \emph {et~al.}(2001)\citenamefont {Cron},
  \citenamefont {Goffman}, \citenamefont {Esteve},\ and\ \citenamefont
  {Urbina}}]{Cron2001}%
  \BibitemOpen
  \bibfield  {author} {\bibinfo {author} {\bibfnamefont {R.}~\bibnamefont
  {Cron}}, \bibinfo {author} {\bibfnamefont {M.~F.}\ \bibnamefont {Goffman}},
  \bibinfo {author} {\bibfnamefont {D.}~\bibnamefont {Esteve}},\ and\ \bibinfo
  {author} {\bibfnamefont {C.}~\bibnamefont {Urbina}},\ }\href
  {https://doi.org/10.1103/PhysRevLett.86.4104} {\bibfield  {journal} {\bibinfo
   {journal} {Phys. Rev. Lett.}\ }\textbf {\bibinfo {volume} {86}},\ \bibinfo
  {pages} {4104} (\bibinfo {year} {2001})}\BibitemShut {NoStop}%
\bibitem [{\citenamefont {Oberholzer}\ \emph {et~al.}(2001)\citenamefont
  {Oberholzer}, \citenamefont {Sukhorukov}, \citenamefont {Strunk},
  \citenamefont {Sch\"onenberger}, \citenamefont {Heinzel},\ and\ \citenamefont
  {Holland}}]{Oberholzer2001}%
  \BibitemOpen
  \bibfield  {author} {\bibinfo {author} {\bibfnamefont {S.}~\bibnamefont
  {Oberholzer}}, \bibinfo {author} {\bibfnamefont {E.~V.}\ \bibnamefont
  {Sukhorukov}}, \bibinfo {author} {\bibfnamefont {C.}~\bibnamefont {Strunk}},
  \bibinfo {author} {\bibfnamefont {C.}~\bibnamefont {Sch\"onenberger}},
  \bibinfo {author} {\bibfnamefont {T.}~\bibnamefont {Heinzel}},\ and\ \bibinfo
  {author} {\bibfnamefont {M.}~\bibnamefont {Holland}},\ }\href
  {https://doi.org/10.1103/PhysRevLett.86.2114} {\bibfield  {journal} {\bibinfo
   {journal} {Phys. Rev. Lett.}\ }\textbf {\bibinfo {volume} {86}},\ \bibinfo
  {pages} {2114} (\bibinfo {year} {2001})}\BibitemShut {NoStop}%
\bibitem [{\citenamefont {Onac}\ \emph {et~al.}(2006)\citenamefont {Onac},
  \citenamefont {Balestro}, \citenamefont {Trauzettel}, \citenamefont
  {Lodewijk},\ and\ \citenamefont {Kouwenhoven}}]{Onac2006}%
  \BibitemOpen
  \bibfield  {author} {\bibinfo {author} {\bibfnamefont {E.}~\bibnamefont
  {Onac}}, \bibinfo {author} {\bibfnamefont {F.}~\bibnamefont {Balestro}},
  \bibinfo {author} {\bibfnamefont {B.}~\bibnamefont {Trauzettel}}, \bibinfo
  {author} {\bibfnamefont {C.~F.~J.}\ \bibnamefont {Lodewijk}},\ and\ \bibinfo
  {author} {\bibfnamefont {L.~P.}\ \bibnamefont {Kouwenhoven}},\ }\href
  {https://doi.org/10.1103/PhysRevLett.96.026803} {\bibfield  {journal}
  {\bibinfo  {journal} {Phys. Rev. Lett.}\ }\textbf {\bibinfo {volume} {96}},\
  \bibinfo {pages} {026803} (\bibinfo {year} {2006})}\BibitemShut {NoStop}%
\bibitem [{\citenamefont {Altimiras}\ \emph {et~al.}(2014)\citenamefont
  {Altimiras}, \citenamefont {Parlavecchio}, \citenamefont {Joyez},
  \citenamefont {Vion}, \citenamefont {Roche}, \citenamefont {Esteve},\ and\
  \citenamefont {Portier}}]{Altimiras2014}%
  \BibitemOpen
  \bibfield  {author} {\bibinfo {author} {\bibfnamefont {C.}~\bibnamefont
  {Altimiras}}, \bibinfo {author} {\bibfnamefont {O.}~\bibnamefont
  {Parlavecchio}}, \bibinfo {author} {\bibfnamefont {P.}~\bibnamefont {Joyez}},
  \bibinfo {author} {\bibfnamefont {D.}~\bibnamefont {Vion}}, \bibinfo {author}
  {\bibfnamefont {P.}~\bibnamefont {Roche}}, \bibinfo {author} {\bibfnamefont
  {D.}~\bibnamefont {Esteve}},\ and\ \bibinfo {author} {\bibfnamefont
  {F.}~\bibnamefont {Portier}},\ }\href
  {https://doi.org/10.1103/PhysRevLett.112.236803} {\bibfield  {journal}
  {\bibinfo  {journal} {Phys. Rev. Lett.}\ }\textbf {\bibinfo {volume} {112}},\
  \bibinfo {pages} {236803} (\bibinfo {year} {2014})}\BibitemShut {NoStop}%
\bibitem [{\citenamefont {Ronen}\ \emph {et~al.}(2016)\citenamefont {Ronen},
  \citenamefont {Cohen}, \citenamefont {Kang}, \citenamefont {Haim},
  \citenamefont {Rieder}, \citenamefont {Heiblum}, \citenamefont {Mahalu},\
  and\ \citenamefont {Shtrikman}}]{Ronen2015}%
  \BibitemOpen
  \bibfield  {author} {\bibinfo {author} {\bibfnamefont {Y.}~\bibnamefont
  {Ronen}}, \bibinfo {author} {\bibfnamefont {Y.}~\bibnamefont {Cohen}},
  \bibinfo {author} {\bibfnamefont {J.-H.}\ \bibnamefont {Kang}}, \bibinfo
  {author} {\bibfnamefont {A.}~\bibnamefont {Haim}}, \bibinfo {author}
  {\bibfnamefont {M.-T.}\ \bibnamefont {Rieder}}, \bibinfo {author}
  {\bibfnamefont {M.}~\bibnamefont {Heiblum}}, \bibinfo {author} {\bibfnamefont
  {D.}~\bibnamefont {Mahalu}},\ and\ \bibinfo {author} {\bibfnamefont
  {H.}~\bibnamefont {Shtrikman}},\ }\href
  {https://doi.org/10.1073/pnas.1515173113} {\bibfield  {journal} {\bibinfo
  {journal} {Proceedings of the National Academy of Sciences}\ }\textbf
  {\bibinfo {volume} {113}},\ \bibinfo {pages} {1743} (\bibinfo {year}
  {2016})},\ \Eprint
  {https://arxiv.org/abs/https://www.pnas.org/content/113/7/1743.full.pdf}
  {https://www.pnas.org/content/113/7/1743.full.pdf} \BibitemShut {NoStop}%
\bibitem [{\citenamefont {van~den Brom}\ and\ \citenamefont {van
  Ruitenbeek}(1999)}]{Brom1999}%
  \BibitemOpen
  \bibfield  {author} {\bibinfo {author} {\bibfnamefont {H.~E.}\ \bibnamefont
  {van~den Brom}}\ and\ \bibinfo {author} {\bibfnamefont {J.~M.}\ \bibnamefont
  {van Ruitenbeek}},\ }\href {https://doi.org/10.1103/PhysRevLett.82.1526}
  {\bibfield  {journal} {\bibinfo  {journal} {Phys. Rev. Lett.}\ }\textbf
  {\bibinfo {volume} {82}},\ \bibinfo {pages} {1526} (\bibinfo {year}
  {1999})}\BibitemShut {NoStop}%
\bibitem [{\citenamefont {Djukic}\ and\ \citenamefont
  {Van~Ruitenbeek}(2006)}]{djukic2006}%
  \BibitemOpen
  \bibfield  {author} {\bibinfo {author} {\bibfnamefont {D.}~\bibnamefont
  {Djukic}}\ and\ \bibinfo {author} {\bibfnamefont {J.}~\bibnamefont
  {Van~Ruitenbeek}},\ }\href
  {https://doi.org/https://doi.org/10.1021/nl060116e} {\bibfield  {journal}
  {\bibinfo  {journal} {Nano letters}\ }\textbf {\bibinfo {volume} {6}},\
  \bibinfo {pages} {789} (\bibinfo {year} {2006})}\BibitemShut {NoStop}%
\bibitem [{\citenamefont {Kumar}\ \emph {et~al.}(2012)\citenamefont {Kumar},
  \citenamefont {Avriller}, \citenamefont {Yeyati},\ and\ \citenamefont {van
  Ruitenbeek}}]{Kumar2012}%
  \BibitemOpen
  \bibfield  {author} {\bibinfo {author} {\bibfnamefont {M.}~\bibnamefont
  {Kumar}}, \bibinfo {author} {\bibfnamefont {R.}~\bibnamefont {Avriller}},
  \bibinfo {author} {\bibfnamefont {A.~L.}\ \bibnamefont {Yeyati}},\ and\
  \bibinfo {author} {\bibfnamefont {J.~M.}\ \bibnamefont {van Ruitenbeek}},\
  }\href {https://doi.org/10.1103/PhysRevLett.108.146602} {\bibfield  {journal}
  {\bibinfo  {journal} {Phys. Rev. Lett.}\ }\textbf {\bibinfo {volume} {108}},\
  \bibinfo {pages} {146602} (\bibinfo {year} {2012})}\BibitemShut {NoStop}%
\bibitem [{\citenamefont {Kumar}\ \emph {et~al.}(2013)\citenamefont {Kumar},
  \citenamefont {Tal}, \citenamefont {Smit}, \citenamefont {Smogunov},
  \citenamefont {Tosatti},\ and\ \citenamefont {van Ruitenbeek}}]{Kumar2013}%
  \BibitemOpen
  \bibfield  {author} {\bibinfo {author} {\bibfnamefont {M.}~\bibnamefont
  {Kumar}}, \bibinfo {author} {\bibfnamefont {O.}~\bibnamefont {Tal}}, \bibinfo
  {author} {\bibfnamefont {R.~H.~M.}\ \bibnamefont {Smit}}, \bibinfo {author}
  {\bibfnamefont {A.}~\bibnamefont {Smogunov}}, \bibinfo {author}
  {\bibfnamefont {E.}~\bibnamefont {Tosatti}},\ and\ \bibinfo {author}
  {\bibfnamefont {J.~M.}\ \bibnamefont {van Ruitenbeek}},\ }\href
  {https://doi.org/10.1103/PhysRevB.88.245431} {\bibfield  {journal} {\bibinfo
  {journal} {Phys. Rev. B}\ }\textbf {\bibinfo {volume} {88}},\ \bibinfo
  {pages} {245431} (\bibinfo {year} {2013})}\BibitemShut {NoStop}%
\bibitem [{\citenamefont {Karimi}\ \emph {et~al.}(2016)\citenamefont {Karimi},
  \citenamefont {Bahoosh}, \citenamefont {Herz}, \citenamefont {Hayakawa},
  \citenamefont {Pauly},\ and\ \citenamefont {Scheer}}]{Karimi2016}%
  \BibitemOpen
  \bibfield  {author} {\bibinfo {author} {\bibfnamefont {M.~A.}\ \bibnamefont
  {Karimi}}, \bibinfo {author} {\bibfnamefont {S.~G.}\ \bibnamefont {Bahoosh}},
  \bibinfo {author} {\bibfnamefont {M.}~\bibnamefont {Herz}}, \bibinfo {author}
  {\bibfnamefont {R.}~\bibnamefont {Hayakawa}}, \bibinfo {author}
  {\bibfnamefont {F.}~\bibnamefont {Pauly}},\ and\ \bibinfo {author}
  {\bibfnamefont {E.}~\bibnamefont {Scheer}},\ }\href
  {https://doi.org/10.1021/acs.nanolett.5b04848} {\bibfield  {journal}
  {\bibinfo  {journal} {Nano Letters}\ }\textbf {\bibinfo {volume} {16}},\
  \bibinfo {pages} {1803} (\bibinfo {year} {2016})},\ \bibinfo {note} {pMID:
  26859711},\ \Eprint
  {https://arxiv.org/abs/https://doi.org/10.1021/acs.nanolett.5b04848}
  {https://doi.org/10.1021/acs.nanolett.5b04848} \BibitemShut {NoStop}%
\bibitem [{\citenamefont {Tewari}\ \emph {et~al.}(2017)\citenamefont {Tewari},
  \citenamefont {Sabater}, \citenamefont {Kumar}, \citenamefont {Stahl},
  \citenamefont {Crama},\ and\ \citenamefont {van Ruitenbeek}}]{Tewari2017}%
  \BibitemOpen
  \bibfield  {author} {\bibinfo {author} {\bibfnamefont {S.}~\bibnamefont
  {Tewari}}, \bibinfo {author} {\bibfnamefont {C.}~\bibnamefont {Sabater}},
  \bibinfo {author} {\bibfnamefont {M.}~\bibnamefont {Kumar}}, \bibinfo
  {author} {\bibfnamefont {S.}~\bibnamefont {Stahl}}, \bibinfo {author}
  {\bibfnamefont {B.}~\bibnamefont {Crama}},\ and\ \bibinfo {author}
  {\bibfnamefont {J.~M.}\ \bibnamefont {van Ruitenbeek}},\ }\href
  {https://doi.org/10.1063/1.5003391} {\bibfield  {journal} {\bibinfo
  {journal} {Review of Scientific Instruments}\ }\textbf {\bibinfo {volume}
  {88}},\ \bibinfo {pages} {093903} (\bibinfo {year} {2017})},\ \Eprint
  {https://arxiv.org/abs/https://doi.org/10.1063/1.5003391}
  {https://doi.org/10.1063/1.5003391} \BibitemShut {NoStop}%
\bibitem [{\citenamefont {Tewari}\ \emph {et~al.}(2019)\citenamefont {Tewari},
  \citenamefont {Sabater},\ and\ \citenamefont {van Ruitenbeek}}]{Tewari2019}%
  \BibitemOpen
  \bibfield  {author} {\bibinfo {author} {\bibfnamefont {S.}~\bibnamefont
  {Tewari}}, \bibinfo {author} {\bibfnamefont {C.}~\bibnamefont {Sabater}},\
  and\ \bibinfo {author} {\bibfnamefont {J.}~\bibnamefont {van Ruitenbeek}},\
  }\href {https://doi.org/10.1039/C9NR05774A} {\bibfield  {journal} {\bibinfo
  {journal} {Nanoscale}\ }\textbf {\bibinfo {volume} {11}},\ \bibinfo {pages}
  {19462} (\bibinfo {year} {2019})}\BibitemShut {NoStop}%
\bibitem [{\citenamefont {Birk}\ \emph {et~al.}(1995)\citenamefont {Birk},
  \citenamefont {de~Jong},\ and\ \citenamefont {Sch\"onenberger}}]{Birk1995}%
  \BibitemOpen
  \bibfield  {author} {\bibinfo {author} {\bibfnamefont {H.}~\bibnamefont
  {Birk}}, \bibinfo {author} {\bibfnamefont {M.~J.~M.}\ \bibnamefont
  {de~Jong}},\ and\ \bibinfo {author} {\bibfnamefont {C.}~\bibnamefont
  {Sch\"onenberger}},\ }\href {https://doi.org/10.1103/PhysRevLett.75.1610}
  {\bibfield  {journal} {\bibinfo  {journal} {Phys. Rev. Lett.}\ }\textbf
  {\bibinfo {volume} {75}},\ \bibinfo {pages} {1610} (\bibinfo {year}
  {1995})}\BibitemShut {NoStop}%
\bibitem [{\citenamefont {Kemiktarak}\ \emph {et~al.}(2007)\citenamefont
  {Kemiktarak}, \citenamefont {Ndukum}, \citenamefont {Schwab},\ and\
  \citenamefont {Ekinci}}]{kemiktarak2007}%
  \BibitemOpen
  \bibfield  {author} {\bibinfo {author} {\bibfnamefont {U.}~\bibnamefont
  {Kemiktarak}}, \bibinfo {author} {\bibfnamefont {T.}~\bibnamefont {Ndukum}},
  \bibinfo {author} {\bibfnamefont {K.}~\bibnamefont {Schwab}},\ and\ \bibinfo
  {author} {\bibfnamefont {K.}~\bibnamefont {Ekinci}},\ }\href
  {https://doi.org/https://doi.org/10.1038/nature06238} {\bibfield  {journal}
  {\bibinfo  {journal} {Nature}\ }\textbf {\bibinfo {volume} {450}},\ \bibinfo
  {pages} {85} (\bibinfo {year} {2007})}\BibitemShut {NoStop}%
\bibitem [{\citenamefont {Herz}\ \emph {et~al.}(2013)\citenamefont {Herz},
  \citenamefont {Bouvron}, \citenamefont {Ćavar}, \citenamefont {Fonin},
  \citenamefont {Belzig},\ and\ \citenamefont {Scheer}}]{Herz2013}%
  \BibitemOpen
  \bibfield  {author} {\bibinfo {author} {\bibfnamefont {M.}~\bibnamefont
  {Herz}}, \bibinfo {author} {\bibfnamefont {S.}~\bibnamefont {Bouvron}},
  \bibinfo {author} {\bibfnamefont {E.}~\bibnamefont {Ćavar}}, \bibinfo
  {author} {\bibfnamefont {M.}~\bibnamefont {Fonin}}, \bibinfo {author}
  {\bibfnamefont {W.}~\bibnamefont {Belzig}},\ and\ \bibinfo {author}
  {\bibfnamefont {E.}~\bibnamefont {Scheer}},\ }\href
  {https://doi.org/10.1039/C3NR02216A} {\bibfield  {journal} {\bibinfo
  {journal} {Nanoscale}\ }\textbf {\bibinfo {volume} {5}},\ \bibinfo {pages}
  {9978} (\bibinfo {year} {2013})}\BibitemShut {NoStop}%
\bibitem [{\citenamefont {Bastiaans}\ \emph {et~al.}(2018)\citenamefont
  {Bastiaans}, \citenamefont {Benschop}, \citenamefont {Chatzopoulos},
  \citenamefont {Cho}, \citenamefont {Dong}, \citenamefont {Jin},\ and\
  \citenamefont {Allan}}]{Bastiaans2018}%
  \BibitemOpen
  \bibfield  {author} {\bibinfo {author} {\bibfnamefont {K.~M.}\ \bibnamefont
  {Bastiaans}}, \bibinfo {author} {\bibfnamefont {T.}~\bibnamefont {Benschop}},
  \bibinfo {author} {\bibfnamefont {D.}~\bibnamefont {Chatzopoulos}}, \bibinfo
  {author} {\bibfnamefont {D.}~\bibnamefont {Cho}}, \bibinfo {author}
  {\bibfnamefont {Q.}~\bibnamefont {Dong}}, \bibinfo {author} {\bibfnamefont
  {Y.}~\bibnamefont {Jin}},\ and\ \bibinfo {author} {\bibfnamefont {M.~P.}\
  \bibnamefont {Allan}},\ }\href {https://doi.org/10.1063/1.5043267} {\bibfield
   {journal} {\bibinfo  {journal} {Review of Scientific Instruments}\ }\textbf
  {\bibinfo {volume} {89}},\ \bibinfo {pages} {093709} (\bibinfo {year}
  {2018})},\ \Eprint {https://arxiv.org/abs/https://doi.org/10.1063/1.5043267}
  {https://doi.org/10.1063/1.5043267} \BibitemShut {NoStop}%
\bibitem [{\citenamefont {Massee}\ \emph {et~al.}(2018)\citenamefont {Massee},
  \citenamefont {Dong}, \citenamefont {Cavanna}, \citenamefont {Jin},\ and\
  \citenamefont {Aprili}}]{Massee2018}%
  \BibitemOpen
  \bibfield  {author} {\bibinfo {author} {\bibfnamefont {F.}~\bibnamefont
  {Massee}}, \bibinfo {author} {\bibfnamefont {Q.}~\bibnamefont {Dong}},
  \bibinfo {author} {\bibfnamefont {A.}~\bibnamefont {Cavanna}}, \bibinfo
  {author} {\bibfnamefont {Y.}~\bibnamefont {Jin}},\ and\ \bibinfo {author}
  {\bibfnamefont {M.}~\bibnamefont {Aprili}},\ }\href
  {https://doi.org/10.1063/1.5043261} {\bibfield  {journal} {\bibinfo
  {journal} {Review of Scientific Instruments}\ }\textbf {\bibinfo {volume}
  {89}},\ \bibinfo {pages} {093708} (\bibinfo {year} {2018})},\ \Eprint
  {https://arxiv.org/abs/https://doi.org/10.1063/1.5043261}
  {https://doi.org/10.1063/1.5043261} \BibitemShut {NoStop}%
\bibitem [{\citenamefont {Mohr}\ \emph {et~al.}(2019)\citenamefont {Mohr},
  \citenamefont {Jasper-Toennies}, \citenamefont {Weismann}, \citenamefont
  {Frederiksen}, \citenamefont {Garcia-Lekue}, \citenamefont {Ulrich},
  \citenamefont {Herges},\ and\ \citenamefont {Berndt}}]{Mohr2019}%
  \BibitemOpen
  \bibfield  {author} {\bibinfo {author} {\bibfnamefont {M.}~\bibnamefont
  {Mohr}}, \bibinfo {author} {\bibfnamefont {T.}~\bibnamefont
  {Jasper-Toennies}}, \bibinfo {author} {\bibfnamefont {A.}~\bibnamefont
  {Weismann}}, \bibinfo {author} {\bibfnamefont {T.}~\bibnamefont
  {Frederiksen}}, \bibinfo {author} {\bibfnamefont {A.}~\bibnamefont
  {Garcia-Lekue}}, \bibinfo {author} {\bibfnamefont {S.}~\bibnamefont
  {Ulrich}}, \bibinfo {author} {\bibfnamefont {R.}~\bibnamefont {Herges}},\
  and\ \bibinfo {author} {\bibfnamefont {R.}~\bibnamefont {Berndt}},\ }\href
  {https://doi.org/10.1103/PhysRevB.99.245417} {\bibfield  {journal} {\bibinfo
  {journal} {Phys. Rev. B}\ }\textbf {\bibinfo {volume} {99}},\ \bibinfo
  {pages} {245417} (\bibinfo {year} {2019})}\BibitemShut {NoStop}%
\bibitem [{\citenamefont {Mohr}\ \emph {et~al.}(2021)\citenamefont {Mohr},
  \citenamefont {Weismann}, \citenamefont {Li}, \citenamefont {Brandbyge},\
  and\ \citenamefont {Berndt}}]{Mohr2021}%
  \BibitemOpen
  \bibfield  {author} {\bibinfo {author} {\bibfnamefont {M.}~\bibnamefont
  {Mohr}}, \bibinfo {author} {\bibfnamefont {A.}~\bibnamefont {Weismann}},
  \bibinfo {author} {\bibfnamefont {D.}~\bibnamefont {Li}}, \bibinfo {author}
  {\bibfnamefont {M.}~\bibnamefont {Brandbyge}},\ and\ \bibinfo {author}
  {\bibfnamefont {R.}~\bibnamefont {Berndt}},\ }\href
  {https://doi.org/10.1103/PhysRevB.104.115431} {\bibfield  {journal} {\bibinfo
   {journal} {Phys. Rev. B}\ }\textbf {\bibinfo {volume} {104}},\ \bibinfo
  {pages} {115431} (\bibinfo {year} {2021})}\BibitemShut {NoStop}%
\bibitem [{\citenamefont {Thupakula}\ \emph {et~al.}(2021)\citenamefont
  {Thupakula}, \citenamefont {Perrin}, \citenamefont {Palacio-Morales},
  \citenamefont {Cario}, \citenamefont {Aprili}, \citenamefont {Simon},\ and\
  \citenamefont {Massee}}]{thupakula2021}%
  \BibitemOpen
  \bibfield  {author} {\bibinfo {author} {\bibfnamefont {U.}~\bibnamefont
  {Thupakula}}, \bibinfo {author} {\bibfnamefont {V.}~\bibnamefont {Perrin}},
  \bibinfo {author} {\bibfnamefont {A.}~\bibnamefont {Palacio-Morales}},
  \bibinfo {author} {\bibfnamefont {L.}~\bibnamefont {Cario}}, \bibinfo
  {author} {\bibfnamefont {M.}~\bibnamefont {Aprili}}, \bibinfo {author}
  {\bibfnamefont {P.}~\bibnamefont {Simon}},\ and\ \bibinfo {author}
  {\bibfnamefont {F.}~\bibnamefont {Massee}},\ }\href@noop {} {\bibinfo {title}
  {Coherent and incoherent tunneling into ysr states revealed by atomic scale
  shot-noise spectroscopy}} (\bibinfo {year} {2021}),\ \Eprint
  {https://arxiv.org/abs/2111.04749} {arXiv:2111.04749 [cond-mat.supr-con]}
  \BibitemShut {NoStop}%
\bibitem [{\citenamefont {Scheer}\ \emph {et~al.}(1998)\citenamefont {Scheer},
  \citenamefont {Agra{\"\i}t}, \citenamefont {Cuevas}, \citenamefont {Yeyati},
  \citenamefont {Ludoph}, \citenamefont {Mart{\'\i}n-Rodero}, \citenamefont
  {Bollinger}, \citenamefont {van Ruitenbeek},\ and\ \citenamefont
  {Urbina}}]{scheer1998}%
  \BibitemOpen
  \bibfield  {author} {\bibinfo {author} {\bibfnamefont {E.}~\bibnamefont
  {Scheer}}, \bibinfo {author} {\bibfnamefont {N.}~\bibnamefont {Agra{\"\i}t}},
  \bibinfo {author} {\bibfnamefont {J.~C.}\ \bibnamefont {Cuevas}}, \bibinfo
  {author} {\bibfnamefont {A.~L.}\ \bibnamefont {Yeyati}}, \bibinfo {author}
  {\bibfnamefont {B.}~\bibnamefont {Ludoph}}, \bibinfo {author} {\bibfnamefont
  {A.}~\bibnamefont {Mart{\'\i}n-Rodero}}, \bibinfo {author} {\bibfnamefont
  {G.~R.}\ \bibnamefont {Bollinger}}, \bibinfo {author} {\bibfnamefont {J.~M.}\
  \bibnamefont {van Ruitenbeek}},\ and\ \bibinfo {author} {\bibfnamefont
  {C.}~\bibnamefont {Urbina}},\ }\href
  {https://doi.org/https://doi.org/10.1038/28112} {\bibfield  {journal}
  {\bibinfo  {journal} {Nature}\ }\textbf {\bibinfo {volume} {394}},\ \bibinfo
  {pages} {154} (\bibinfo {year} {1998})}\BibitemShut {NoStop}%
\bibitem [{\citenamefont {W\"oll}\ \emph {et~al.}(1989)\citenamefont {W\"oll},
  \citenamefont {Chiang}, \citenamefont {Wilson},\ and\ \citenamefont
  {Lippel}}]{Woll1989}%
  \BibitemOpen
  \bibfield  {author} {\bibinfo {author} {\bibfnamefont {C.}~\bibnamefont
  {W\"oll}}, \bibinfo {author} {\bibfnamefont {S.}~\bibnamefont {Chiang}},
  \bibinfo {author} {\bibfnamefont {R.~J.}\ \bibnamefont {Wilson}},\ and\
  \bibinfo {author} {\bibfnamefont {P.~H.}\ \bibnamefont {Lippel}},\ }\href
  {https://doi.org/10.1103/PhysRevB.39.7988} {\bibfield  {journal} {\bibinfo
  {journal} {Phys. Rev. B}\ }\textbf {\bibinfo {volume} {39}},\ \bibinfo
  {pages} {7988} (\bibinfo {year} {1989})}\BibitemShut {NoStop}%
\end{thebibliography}%

\end{document}